\documentstyle[12pt]{article}
\begin{document}
\def\beq{\begin{equation}}
\def\eeq{\end{equation}}
\def\bea{\begin{eqnarray}}
\def\eea{\end{eqnarray}}
\def\ve{\vert}
\def\vel{\left|}
\def\ver{\right|}
\def\nnb{\nonumber}
\def\ga{\left(}
\def\dr{\right)}
\def\aga{\left\{}
\def\adr{\right\}}
\def\rar{\rightarrow}
\def\nnb{\nonumber}
\def\la{\langle}
\def\ra{\rangle}
\def\ba{\begin{array}}
\def\ea{\end{array}}
\def\tep{$B \rar K^* l^+ l^-$}
\def\tepm{$B \rar K^* \mu^+ \mu^-$} 
\def\tept{$B \rar K^* \tau^+ \tau^-$}

\title{ {\small { \bf RARE $B \rar K^* l^+ l^-$ DECAY
                      IN LIGHT CONE QCD} } }

\author{ {\small T. M. AL\.{I}EV \thanks
{e-mail:taliev@rorqual.cc.metu.edu.tr}\,\,,
A. \"{O}ZP\.{I}NEC\.{I} \thanks 
{e-mail:e100690@orca.cc.metu.edu.tr}\,\, and \,\,
M. SAVCI }\\ 
{\small Physics Department, Middle East Technical University} \\
{\small 06531 Ankara, Turkey} }

\begin{titlepage}
\maketitle
\thispagestyle{empty}

\begin{abstract}
\baselineskip  0.7cm

We investigate the transition formfactors for the 
$B \rar K^* l^+ l^-$($l = \mu,~ \tau$)
decay in the light cone QCD. It is found that the light cone and 3-point QCD
sum rules analyses for some of the formfactors for the decay 
$B \rar K^* l^+ l^-$ lead to
absolutely different $q^2$ dependence. The invariant dilepton mass distributions for
the $B \rar K^* \mu^+ \mu^-$ and $B \rar K^* \tau^+ \tau^-$ 
decays and final lepton longitudinal polarization asymmetry, which
includes both short and long-distance contributions, are also calculated.

\end{abstract}

\vspace{1cm}
~~~PACS numbers: 13.20.He, 11.55.Hx, 12.38.Cy
\end{titlepage}

\setcounter{page}{1}
\section{Introduction}
Experimental observation \cite{R1} of the inclusive
and exclusive radiative decays $B \rar X_s \gamma$ and $B \rar K^* \gamma$
stimulated the study of rare $B$ decays on a new footing.  These flavor
changing neutral current (FCNC) $b \rar s$ transitions in the SM do not
occur at tree level and appear only at loop  level.  Therefore the
study of these rare $B$-meson decays can provide a means of testing the
detailed structure of the SM at the loop level. These decays are also very
useful for extracting the values of the Cabibbo-Kobayashi-Maskava (CKM)
matrix elements \cite{R2}, as well as for establishing new  physics beyond
the Standard Model \cite{R3}.

Currently, the main interest on rare $B$-meson decays is focused on
decays for which the SM predicts large branching ratios  and can be potentially
measurable in the near future. The rare $B \rar K l^+ l^-$ and 
$B \rar K^* l^+ l^-$
decays are such decays. The experimental situation for these decays is very
promising \cite{R4}, with $e^+ e^-$ and hadron colliders focusing only on
the observation of exclusive modes with $l = e,~\mu$ and $\tau$ final states,
respectively.  At quark level the process $b \rar s l^+ l^-$ takes place
via electromagnetic and Z penguins and W box diagrams and are described by
three independent Wilson coefficients $C_7,~C_9$ and $C_{10}$.
Investigations
allow us to study different structures, described by the above mentioned
Wilson coefficients. In the SM, the measurement of the forward-backward
asymmetry and invariant dilepton mass distribution in 
$b \rar q l^+ l^-~(q = s,~d)$
provide information on the short distance contributions dominated by the
top quark loops and are essential in separating the short distance FCNC
process from the contributing long distance effects \cite{R5} and also are very
sensitive to the contributions from new physics \cite{R6}. Recently it has
been emphasized by Hewett \cite{R7} that the longitudinal
lepton polarization, which is another parity violating observable, is also
an important asymmetry and that the lepton polarization in $b \rar s l^+ l^-$
will be measurable with the high statistics available at the B-factories  
currently under construction.
However,
in calculating the Branching ratios and other observables in hadron level,
i.e. for $B \rar K^* l^+ l^-$ decay, we have the problem
of computing the matrix element of the
effective Hamiltonian, ${\cal H}_{eff}$, between the states $B$ and $K^*$.  But
this problem is related to the non-perturbative sector of QCD.

These matrix elements, in the framework of different approaches such as
chiral theory \cite{R8}, three point QCD sum rules method \cite{R9},
relativistic quark model by the light-front formalism \cite{R10,R11}, have
been investigated.
The aim of this work is the calculation of these matrix elements in light
cone QCD sum rules method and to study the lepton polarization asymmetry for
the exclusive $B \rar K^* l^+ l^-$ decays.

The effective Hamiltonian for the $b \rar s l^+ l^-$ decay, including
QCD corrections [12-14] can be written as

\bea
{\cal H}_{eff} = \frac{4 G_F}{\sqrt 2} V_{tb} V^*_{ts} \sum_{i=1}^{10}
C_i( \mu ) O_i( \mu )
\eea
which is evolved from the electroweak scale down to $\mu \sim m_b$ by the
renormalization group equations. Here $V_{ij}$ represent the relevant CKM
matrix elements, and $O_i$ are a complete set of renormalized dimension 5
and 6 operators involving light fields which govern the $b \rar s$
transitions and $C_i ( \mu )$ are the Wilson coefficients for the corresponding
operators. The explicit forms of  $C_i ( \mu )$ and $O_i ( \mu )$ can be
found in [12-14]. For $b \rar s l^+ l^-$ decay, this
effective Hamiltonian leads to the matrix element

\bea
{\cal M} = \frac{G_F \alpha}{\sqrt 2 \pi} V_{tb} V^*_{ts} \left[ C_9^{eff}
\bar s_L \gamma_\mu b_L~ \bar l \gamma^\mu l + C_{10} \bar s_L \gamma_\mu b_L~
\bar l \gamma^\mu \gamma_5 l - 2 \frac{C_7}{q^2}\bar s i \sigma_{\mu \nu} q^\nu
(m_b R + m_s L) b~ \bar l \gamma^\mu l \right]
\eea
where $q^2$ is the invariant dilepton mass, and
$L(R)=\left[1-(+)\gamma_5\right]/2$ are the projection operators.
  The coefficient $C_9^{eff} (
\mu,~q^2) \equiv C_9( \mu) + Y ( \mu,~q^2)$,
 where the function $Y$ contains the
contributions from the one loop matrix element of the four-quark
operators and can be found in [12-14]. Note that the function $Y
( \mu,~q^2)$ contains both real and imaginary parts (the imaginary part
arises when the c-quark in the loop is on the mass shell).

The $B \rar K^* l^+ l^-$ decay also receives large long 
distance contributions from the
cascade process $B \rar K^* \psi^{(_{'})}\rar K^* l^+ l^-$. These contributions are
taken into account by introducing a Breit-Wigner form of the resonance
propogator and this procedure leads to an additional contribution to
$C_9^{eff}$ of the form \cite{R15}

\bea
-\, \frac{2 \pi}{\alpha^2} \sum_{V= \psi,~ \psi'} \frac{m_V \Gamma(V \rar l^+
l^-)}{(q^2 - m_V^2) - i m_V \Gamma_V}
\nnb
\eea

As we noted earlier, for the calculation of the branching ratios for the exclusive
$B \rar K^* l^+ l^-$ decays, first of all, we must calculate the matrix elements
$\la K^* \ve \bar s \gamma_\mu (1- \gamma_5) q \ve B \ra$  and
$\la K^* \ve \bar s i \sigma_{\mu \nu} p^{\nu} (1+\gamma_5) q \ve B \ra$.
These matrix elements can be parametrized in terms of the formfactors as follows
(see also \cite{R9}):

\bea
\la K^* ( p, \epsilon)  \vel \bar s \gamma_\mu ( 1-\gamma_5 ) q \ver B ( p+q ) \ra &=&
 -\; \epsilon_{\mu\nu\rho\sigma} \, \epsilon^{*\nu} p^\rho q^\sigma 
\frac{2 V ( q^2 )}{m_B + m_{K^*}} - \nnb \\
&& - \; i \, \epsilon_\mu^* \ga m_B + m_{K^*}\dr A_1 ( q^2 ) + \nnb \\
&&+\; i \, (  \epsilon^* q )  P_\mu \frac{A_2 ( q^2 )}{m_B +
m_{K^*}} + \nnb \\ 
&& +\; i \, \ga \epsilon^* q \dr \frac{2 m_{K^*}}{q^2} 
\left[A_3 ( q^2 ) - A_0 ( q^2 ) \right] q_\mu ~, 
\\ \nnb \\ \nnb \\ \nnb
\la K^* ( p, \epsilon ) \vel \bar s i \sigma_{\mu\nu} q^\nu 
( 1 + \gamma_5 ) q \ver B (  p+q ) \ra &=&
4 \;\epsilon_{\mu\nu\rho\sigma} \, \epsilon^{*\nu} p^\rho q^\sigma
\,  T_1 ( q^2 ) + \nnb \\
&& + \;2\; i  \left[ \epsilon^*_\mu (P q) - 
( \epsilon^* q ) P_\mu \right] T_2 ( q^2 ) + \nnb \\
&& + \;2\; i \, (  \epsilon^* q )\left[ q_\mu - \frac{q^2}{Pq}
P_\mu \right] T_3 ( q^2 )  ~, \nnb \\
\\ \nnb
\eea
where $\epsilon_\mu^*$ is the polarization vector of $K^*$, $p+q$ and
$p$ are the momentum of $B$ and  $K^*$ and  $P_\mu = \ga 2 p + q \dr_\mu$.
The formfactor $A_3 \ga q^2 \dr$ can be written as a linear combination of 
the formfactors $A_1$ and $A_2$ (see \cite{R9}):
\beq
A_3 ( q^2 ) = \frac{m_B + m_{K^*}}{2 m_{K^*}} \, A_1 ( q^2 ) - 
\frac{m_B - m_{K^*}}{2 m_{K^*}} \, A_2 ( q^2 )~,
\eeq
with the condition $A_3(0) = A_0(0)$. In calculating these formfactors
we employ the light cone QCD sum rules.

\section{QCD Sum Rules for Formfactors}

According to the QCD sum rules ideology, in order to calculate the 
formfactors we start by considering the representation of a suitable
correlator function in terms of hadron language  and quark-gluon language.
Equating these representations we get the sum rules. For this purpose we
choose the following correlators.
\bea
\Pi_{\mu}^{( 1 )}(p,q) &=& i \, \int{d^4 x e^{i q x} \la K^*(p) \ve \bar s(x)
\gamma_\mu ( 1-\gamma_5 ) b(x) \bar b(0) i \gamma_5 q(0) \ve 0 \ra} ~,
\\ \nnb \\ 
\Pi_{\mu}^{( 2 )}(p,q) &=& i \, \int{d^4 x e^{i q x} \la K^*(p) \ve \bar s(x)
i \sigma_{\mu\nu} q^\nu ( 1 + \gamma_5 )b(x) \bar b(0) i \gamma_5 q(0) \ve 0 \ra} ~.
\eea
Here the first correlator is relevant for the calculation of the formfactors
$V(q^2),~A_1(q^2),~A_2(q^2)$ and $A_0(q^2)$ and the second one for
$T_1,~T_2$ and $T_3$.

The main task in QCD is the calculation of the correlation functions (6) and
(7). This problem can be solved in the deep Euclidean region, where both
virtualities $q^2$ and $(p+q)^2$ are large and negative. The virtuality of
the heavy quark in the correlators (6) and (7) is large, of order
$m_b^2 - (p+q)^2$, and one can use the perturbative expansion of its
propagator in the external field of slowly varying fluctuations inside the
vector meson. Then, the  leading contribution is
\bea
\Pi_{\mu}^{( 1 )}(p,q) &=& i \, \int{ \frac{d^4 x\, d^4 k}{(2 \pi)^4} \,
\frac{e^{i(q-k)x}}{(m_b^2-k^2)}} 
\la K^* \ve \bar s(x) \gamma_\mu (1-\gamma_5) (\not\!k+m_b)
\gamma_5  q(0) \ve 0 \ra~, \\ \nnb \\ 
\Pi_{\mu}^{( 2 )}(p,q) &=& - \int{ \frac{d^4 x \, d^4 k }{(2 \pi)^4} \,
\frac{e^{i(q-k)x}}{(m_b^2-k^2)}}
\, q^\nu \, \la K^* \ve \bar s(x) \sigma_{\mu\nu}
(1+\gamma_5) (\not\!k+m_b) \gamma_5  q(0) \ve 0 \ra~. 
\eea
It is obvious from the above expressions that the problem is reduced to the
calculation of the matrix
elements of the gauge-invariant non-local operators, sandwiched in between
the vacuum and the meson states. These matrix elements define the vector
meson light cone
wave functions. Following \cite{R16,R17} we define the meson wave functions
as:

\bea
\la 0 \ve \bar q (0) \sigma_{\mu \nu} q (x) \ve K^* (p,\epsilon) \ra &=&
 i \, \ga \epsilon_\mu p_\nu - \epsilon_\nu p_\mu \dr f_{K^*}^{\perp}
\int_0^1{du\, e^{-iu p x}} \phi_\perp (u,\mu^2)~, \\ \nnb \\
\la 0 \ve \bar q (0) \gamma_\mu q(x)  \ve K^* (p,\epsilon) \ra &=&
p_\mu \frac{\epsilon x}{px} f_{K^*} m_{K^*} \int_0^1{du\, e^{-iu p x}}
\phi_\parallel(u,\mu^2) +  \\ \nnb
&& +\; \ga \epsilon_\mu - p_\mu \frac{\epsilon x}{p x} \dr f_{K^*} m_{K^*}
\int_0^1{du \, e^{-iu p x}} g_{\perp}^{(v)} (u,\mu^2)~,  \\ \nnb \\
\la 0 \ve \bar q (0) \gamma_\mu \gamma_5 q(x)  \ve K^* (p,\epsilon) \ra &=& 
- \, \frac{1}{4}\, \epsilon_{\mu \nu \rho \sigma} \epsilon^\nu p^\rho x^\sigma
f_{K^*} m_{K^*} \int_0^1{du \, e^{-iu p x}} g_\perp^{(a)} (u,\mu^2)~.
\eea
The functions $\phi_\perp ( u ,~ \mu^2)$ and $\phi_\parallel(u,\mu^2)$ give
the leading twist distributions in the fraction of total momentum carried by
the quark in the transversaly and longitudinally polarized meson,
respectively.  In \cite{R17} it was shown that

\bea
g_\perp^v ( u) &=& \frac{3}{4} \left[ 1 + (2 u - 1)^2 \right]~,  \\
g_\perp^a ( u) &=& 6 u ( 1 - u) ~,
\eea
which we use in the numerical analysis.
For the explicit form of  $\phi_\perp ( u ,~ \mu^2)$ we shall use the
results of \cite{R17}:
\bea
\phi_\perp ( u ,~ \mu^2) &=& 6 u (1-u) \left\{ 1 + a_1(\mu)(2 u -1) +
a_2(\mu)\left[ (2u-1)^2-\frac{1}{5}\right] +  \right. \nnb \\
&&\left. +\; a_3(\mu)\left[ \frac{7}{3}(2u-1)^3 - (2u-1)
\right]+...\right\}~, \\ \nnb \\
a_n(\mu) &=&
a_n(\mu_0)\left[ \frac{\alpha_s(\mu)}{\alpha_s(\mu_0)}\right]
^{\mbox{\normalsize $\frac{\gamma_n}{b}$}}~.
\eea
Here $b=\frac{11}{3} N_C - \frac{2}{3} n_f$, and
\bea
\gamma_n = C_F \ga 1 + 4 \sum_{j=2}^{n+1} \frac{1}{j} \dr~,
\eea
where $C_F = \frac{N_C^2 - 1}{2 N_C}$.
\\
As in  \cite{R17}, we  will use the following values for the  parameters 
appearing in eqs.(10)-(12) and eq.(15) :
\bea
&&
f_{K^*}^\perp = 210~MeV,~~a_1^{K^*} ( \mu = m_b) = 0.57,~~a_2^{K^*} ( \mu =
m_b) = -1.35 \nnb \\
&&\mbox{and} ~~a_3^{K^*} ( \mu = m_b) = 0.46~, \nnb \\
&&\phi_\parallel( u,~ \mu^2 ) = 6 u (1-u).
\eea
Using eqs.(10-12), we get the following results from eq.(8) and eq.(9)
for the theoretical part 
of the sum rules:

\bea
\Pi_\mu^{(1)} &=& -\;i m_b f_{K^*} m_{K^*} \int_0^1 \frac{du}{\Delta} \left[
\epsilon^*_\mu g_\perp^{(v)}  + 2 ( q \epsilon^*) p_\mu \frac{1}{\Delta}
(\Phi_\parallel - G_\perp^{(v)}) \right] -\nnb  \\
&&-\;\epsilon_{\mu \nu \rho \sigma} \epsilon^{* \nu} p^\rho q^\sigma \left[
\frac{m_b}{2}f_{K^*} m_{K^*} \int_0^1 \frac{du}{\Delta^2} g_\perp^{(a)} +
f_{K^*}^\perp \int_0^1 du \frac{\phi_\perp}{\Delta} \right] - \nnb \\
&& -\; i f_{K^*}^{\perp} \int_0^1 du \frac{\phi_\perp}{\Delta} \left[
\epsilon^*_\mu
(pq + p^2 u) - p_\mu (q \epsilon^*) \right] \\ \nnb \\ \nnb
\eea
\newpage
\bea
\Pi_\mu^{(2)} &=& \epsilon_{\mu\nu\rho\sigma} \epsilon^{* \nu} p^\rho q^\sigma
\left\{m_b f_{K^*}^\perp \int_0^1  \frac{du}{\Delta} \phi_\perp - 
m_{K^*} f_{K^*}\left[ \int_0^1 \frac{du}{\Delta} \ga \Phi_\parallel -
G_\perp^{(v)} \dr  - \right. \right. \nnb \\
&&\left. \left. - \; \int_0^1 \frac{du}{\Delta}u g_\perp^{(v)} -
\int_0^1 \frac{du}{2 \Delta^2}g_\perp^{(a)} \ga \Delta + q^2 + 2 p q u \dr
\right] \right\} + \nnb \\
&&+ \; i \left[ \epsilon^*_\mu (p q) - (q \epsilon^*) p_\mu \right]  \Bigg\{
m_b f_{K^*}^\perp \int_0^1  \frac{du}{\Delta}\phi_\perp + \nnb \\ 
&& + \; m_{K^*} f_{K^*} \int_0^1 \frac{du}{\Delta} \Bigg[ - 
\ga \Phi_\parallel - G_\perp^{(v)} \dr + 
u g_\perp^{(v)} + \frac{g_\perp^{(a)}}{2}
+ \frac{g_\perp^{(a)} u (q p)}{2 \Delta} \Bigg] \Bigg\} + \nnb \\
&& + \;i\, m_{K^*} f_{K^*}\left[ \epsilon^*_\mu (q^2) - (q \epsilon^*) q_\mu
\right]
\int_0^1 \frac{d u}{\Delta} \left[g_\perp^{(v)} - \frac{p^2 u}{2 \Delta}
g_\perp^{(a)} \right] +\nnb \\
&&+ \; 2 im_{K^*} f_{K^*} (q \epsilon^*) \left[ p_\mu (q^2) - (p q) q_\mu
\right]  \int_0^1 \frac{d u}{\Delta^2} \ga  \Phi_\parallel - G_\perp^{(v)} \dr~,
\eea  
where 
\bea  
\Phi_\parallel(u) &=& -\, \int_0^u \phi_\parallel(v) d v~, \nnb \\
G_\perp^{(v)}(u) &=& -\, \int_0^u g_\perp^{(v)}(v) d v~, 
\eea 
and  
\bea 
\Delta &=& m_b^2 - (q + p u)^2~. \nnb
\eea
Let us turn our attention to the physical part of the correlator functions
(6) and (7).
Writing a dispersion relation in the variable $(p + q)^2$, one can separate the
 $B$ meson ground state contribution to the correlator 
functions $\Pi_\mu^{(1)}$ and  $\Pi_\mu^{(2)}$, by inserting a complete set
of states between the currents in (6) and (7) focusing on the term $\ve B
\ra~ \la B \ve$ :

\bea
\Pi_\mu^{(1)} &=& \frac{f_B m_B^2}{m_b \left[ m_B^2 - (q + p)^2
\right]} \, \la K^*(p) \ve \bar s \gamma_\mu (1-\gamma_5) q \ve B(p+q) \ra \\
\Pi_\mu^{(2)} &=& \frac{f_B m_B^2}{m_b \left[ m_B^2 - (q + p)^2 
\right]} \, \la K^*(p) \ve \bar s i \sigma_{\mu\alpha} q^\alpha
(1+\gamma_5) q \ve B(p+q) \ra
\eea
Using the definitions of the formfactors (see eqs.(3) and (4)) in
(22) and (23) and equating these expressions to eqs.(19) and (20), we get
the sum rules for the formfactors. The remaining part of the calculation
follows from the QCD sum rules procedure: perform the Borel transformation
on the variable $(p+q)^2$ and subtract the continuum and higher states
contributions invoking quark-hadron duality. (Details of these procedures
can be found in [17-19]). After this procedure we obtain the following sum
rules for the formfactors: 
\newpage

\bea
V(q^2) &=& \frac{m_B + m_{K^*}}{2}\, \frac{m_b}{f_B m_B^2}
\, e^{\frac{m_B^2}{M^2}}
\int_\delta^1 d u \, exp \ga - \frac{m_b^2 + p^2 u
\bar u - q^2 \bar u}{u M^2} \dr \times \nnb \\ 
&& \times \left\{ m_b f_{K^*} m_{K^*} \frac{g_\perp^{(a)}}{2 u^2 M^2} +
\frac{f_{K^*}^\perp \phi_\perp}{u} \right\}~, \\ \nnb \\ \nnb \\
A_1(q^2) &=& \frac{1}{m_B + m_{K^*}}\, \frac{m_b}{f_B m_B^2}
\, e^{\frac{m_B^2}{M^2}}
\int_\delta^1 d u \, exp \ga - \frac{m_b^2 + p^2 u
\bar u - q^2 \bar u}{u M^2} \dr \times \nnb \\
&& \times \left\{ m_b f_{K^*} m_{K^*} \frac{g_\perp^{(v)}}{u} +
\frac{f_{K^*}^\perp \phi_\perp (m_b^2 - q^2 +p^2 u^2)}{2 u^2}  
\right\}~, \\ \nnb \\ \nnb \\
A_2(q^2) &=& - \, \ga m_B + m_{K^*} \dr \frac{m_b}{f_B m_B^2}
\, e^{\frac{m_B^2}{M^2}}
\int_\delta^1 d u \, exp \ga - \frac{m_b^2 + p^2 u
\bar u - q^2 \bar u}{u M^2} \dr \times \nnb \\
&& \times \left\{ \frac{m_b f_{K^*} m_{K^*}}{u^2 M^2}
\ga \Phi_\parallel - G_\perp^{(v)} \dr - \frac{1}{2} \, f_{K^*}^\perp
\frac{\phi_\perp}{u} \right\}~, \\ \nnb \\ \nnb \\
A_3(q^2) - A_0(q^2) &=& \frac{q^2}{2 m_{K^*}} \, \frac{m_b}{f_B m_{B}^2}
\,e^{\frac{m_B^2}{M^2}}                
\int_\delta^1 d u \, exp \ga - \frac{m_b^2 + p^2 u
\bar u - q^2 \bar u}{u M^2} \dr \times \nnb \\
&& \times \left\{ \frac{m_b f_{K^*} m_{K^*}}{u^2 M^2} \ga \Phi_\parallel -
G_\perp^{(v)} \dr - \frac{1}{2} \, f_{K^*}^\perp
\frac{\phi_\perp}{u} \right\}~.
\eea
From eq.(26) and eq.(27) we get a new relation between formfactors $A_3,~A_0$ and
$A_2$:
\bea  
A_3(q^2) - A_0(q^2) = - \, \frac{A_2(q^2) q^2}{2 m_{K^*} (m_B + m_{K^*})}~.
\eea  
For the formfactors $T_1,~ T_2$, and $T_3$, we get the following sum rules:
\bea
T_1(q^2) &=& \frac{1}{4} \, \frac{m_b}{f_B m_B^2}\, e^{\frac{m_B^2}{M^2}}
\int_\delta^1 \frac{du}{u}\, exp \ga - \frac{m_b^2 + p^2 u \bar u - q^2 \bar u}{u
M^2} \dr \Bigg\{ m_b f_{K^*}^\perp \phi_\perp - \nnb \\
&&-\; f_{K^*} m_{K^*} \Bigg[ \Phi_\parallel  - G_\perp^{(v)} -
u g_\perp^{(v)}  - \frac{g_\perp^{(a)}}{4} - \frac{g_\perp^a(m_b^2 + q^2 -
p^2 u^2)}{4 u M^2} \Bigg] \Bigg\}~, \nnb \\ \nnb \\
\eea

\newpage
\bea
T_2(q^2) &=& \frac{1}{2 (m_{B^*}^2 - m_{K^*}^2)} \, \frac{m_b}{f_B m_B^2}
 \, e^{\frac{m_B^2}{M^2}}
\int_\delta^1 \frac{du}{u}\, exp( - \frac{m_b^2 + p^2 u \bar u - q^2 \bar u}{u
M^2}) \times \nnb \\
&&\times \Bigg\{ f_{K^*} m_{K^*} \left[ g_\perp^{(v)} - \frac{p^2}{2 M^2}
g_\perp^{(a)} \right]q^2 + \frac{m_b f_{K^*}^\perp \phi_\perp}{2 u}( m_b^2 -
q^2 - p^2 u^2)\; + \nnb \\
&&+\; f_{K^*} m_{K^*} \Bigg[ \frac{ (m_b^2 -q^2- p^2 u^2)}{2 u} \times \nnb \\
&&  \times \Bigg( - \left[\Phi_\parallel  - G_\perp^{(v)}\right] +
u g_\perp^{(v)}  + \frac{(m_b^2 -q^2-p^2 u^2) g_\perp^{(a)}}{4 u M^2}
\Bigg) \Bigg] \Bigg\}~, \\ \nnb \\ \nnb \\
T_3(q^2) &=& \frac{1}{4} \, \frac{m_b}{f_B m_B^2} \,e^{\frac{m_B^2}{M^2}}
\int_\delta^1 \frac{du}{u}\, 
exp \ga - \frac{m_b^2 + p^2 u \bar u - q^2 \bar u}{uM^2}\dr
\times \nnb \\
&&\times \Bigg\{ m_{K^*} f_{K^*} \Bigg[ \frac{g_\perp^{(a)}}{4} + \frac{(m_b^2
-q^2-p^2 u^2)}{4 u M^2} g_\perp^{(a)} \Bigg] - \nnb \\
&& - \; 2 m_{K^*} f_{K*} \Bigg[
\frac{g_\perp^{(v)}}{2}(2-u) - \frac{p^2 g_\perp^{(a)}}{2 M^2}
\Bigg]- \nnb \\
&& - \; 2 m_{K^*} f_{K*} \Bigg[ \frac{\Phi_\parallel  -
G_\perp^{(v)}}{u M^2} \Bigg( \frac{m_b^2 - q^2 - p^2 u^2}{u}\, +  \nnb \\
&& + \; q^2 - M^2 +
\frac{u M^2}{2}\Bigg) \Bigg] + m_b f_{K^*}^\perp \phi_\perp \Bigg\}~.
\eea
Using the equation of motion we can relate $T_3$ and $A_3 - A_0$ by:

\bea
T_3(q^2) = m_{K^*} ( m_b - m_s) \frac{A_3(q^2) - A_0(q^2)}{q^2}~.
\eea
Here $M$ is the Borel mass parameter. The lower integration limit $\delta =
\frac{m_b^2-p^2}{s_0 - p^2}$ depends on the effective threshold $s_0$ above
which the contributions from higher states to the dispersion relation (22)
and (23) are cancelled against the corresponding piece in the QCD
representation (19) and (20)
Note that the sum rules for $V(q^2)$ and $A_1(q^2)$ and $T_1(q^2)$ in the
light cone QCD are derived in \cite{R17}. Our results agree with that of 
\cite{R17}. The region of applicability of these sum rules is restricted by
the requirement that the value of $q^2 - m_b^2$ be sufficiently less than
zero.  In order not to introduce an additional scale, we require that $q^2 -
m_b^2 \leq (p+q)^2 - m_b^2$ which translates to the condition that $m_b^2 -
q^2$ is of the order of  the typical Borel parameter $M^2 \sim 5 \div 8~GeV^2$. From
this condition we obtain that the region of applicability of the sum rules is 
$q^2
< 15 \div 17 ~GeV^2$ , which is few $GeV^2$ below the zero recoil point.

\newpage

Finally we calculate the differential decay rate with
longitudinal polarization of the final leptons.  The differential decay rate is
given by:

\bea
\frac{d \Gamma}{d q^2} &=&
\frac{G^2 \alpha^2}{2^{12} \pi^5} \, 
\frac{\vel V_{tb} V_{ts}^* \ver ^2 \sqrt{\lambda} v}{3 m_B}
\Bigg\{
(2 m_l^2 + m_B^2 s)
\Bigg[16 \ga \ve A \ve ^2 + \ve C \ve ^2 \dr
 m_B^4 \lambda  +  \nnb \\
&& + \; 2 \ga \ve B_1 \ve ^2 + \ve D_1 \ve ^2 \dr \frac{\lambda + 12 r s}{r s} 
+ 2 \ga \ve B_2 \ve ^2 + \ve D_2 \ve ^2 \dr \frac{m_B^4 \lambda^2}{r s} -  \nnb \\
&& - \; 4\left[ Re \ga B_1 B_2^* \dr + Re \ga D_1 D_2^* \dr \right]
 \frac{m_B^2 \lambda}{r s} (1 - r -s ) 
\Bigg] + \nnb \\
&& + \; 6 m_l^2 \Bigg[ -\, 16 \ve C \ve ^2 m_B^4 \lambda +
4 Re \ga D_1 D_3^* \dr \frac{m_B^2 \lambda}{r} - \nnb \\  
&&- \;4  Re \ga D_2 D_3^* \dr \frac{m_B^4 (1-r) \lambda}{r} +
2 \ve D_3 \ve ^2 \frac{m_B^4 s \lambda}{r} - 
4 Re \ga D_1 D_2^* \dr \frac{m_B^2 \lambda}{r} - \nnb \\
&& - \; 24 \ve D_1 \ve ^2 + 2 \ve D_2 \ve ^2 \frac{m_B^4 \lambda}{r}
(2  + 2 r - s ) 
\Bigg] - \nnb \\    
&& - \; 4 v \xi \Bigg[
8 Re \ga A C^* \dr \lambda m_B^6 s - 
\left[ Re \ga B_1^* D_2 \dr +Re \ga B_2^* D_1 \dr \right]
\frac{m_B^4 \lambda}{r} (1-r-s) + \nnb \\
&&+ \;  Re \ga B_2^* D_2 \dr \frac{m_B^6 \lambda^2}{r} +    
 \; Re \ga B_1^* D_1 \dr m_B^2 \frac{\lambda + 12 r s}{r}
\Bigg]
\Bigg\}~,
\eea   
where $\lambda   = 1 + r^2 + s^2 - 2 r - 2 s - 2 r s$, $r =
\frac{m_{K^*}^2}{m_B^2}$, $s = \frac{q^2}{m_B^2}$, $\xi$ is the longitudinal
polarization of the final lepton, $m_l$  and $v= \sqrt{1-
\frac{4 m_l^2}{q^2}}$ are its mass and velocity, respectively.
 In eq.(33) $A,~B_1,~B_2,~B_3,~C,~D_1,~D_2$,
and $D_3$ are defined as follows:

\bea
A &=& C_9^{eff} \frac{V}{m_B + m_{K^*}} + 4 C_7 \frac{m_b}{q^2} T_1~, \nnb \\ \nnb \\
B_1 &=& C_9^{eff} (m_B + m_{K^*}) A_1 + 4 C_7 \frac{m_b}{q^2} (m_B^2 -
m_{K^*}^2)~,  \nnb \\ \nnb \\ 
B_2 &=& C_9^{eff} \frac{A_2}{m_B + m_{K^*}} + 4 C_7 \frac{m_b}{q^2} \ga T_2 +
 \frac{q^2}{m_B^2 - m_{K^*}^2} T_3 \dr~,  \nnb \\ \nnb \\
B_3 &=& - C_9^{eff} \frac{2 m_{K^*}}{q^2} ( A_3 - A_0) + 4 C_7 \frac{m_b}{q^2} T_3
~, \nnb \\ \nnb \\
C &=& C_{10} \frac{V}{m_B + m_{K^*}}~,  \nnb \\ \nnb \\
D_1 &=& C_{10} (m_B + m_{K^*}) A_1~,  \nnb \\ \nnb \\
D_2 &=& C_{10} \frac{A_2}{m_B + m_{K^*}}~,  \nnb \\ \nnb \\
D_3 &=& - C_{10} \frac{2 m_{K^*}}{q^2} (A_3 - A_0)~. \nnb
\eea
For the dileptonic decays of the $B$ mesons, the longitudinal polarization
asymmetry, $P_L$, of  
the final  lepton, $l$,  is  defined as

\bea
P_L ( q^2) = 
\frac{
\mbox{\Large $\frac{d \Gamma}{dq^2}$} ( \xi = -1 ) - 
\mbox{\Large $\frac{d \Gamma}{dq^2}$} ( \xi = 1 )         
}
{
\mbox{\Large $\frac{d \Gamma}{dq^2}$} ( \xi = -1) +
\mbox{\Large $\frac{d \Gamma}{dq^2}$} ( \xi = 1 )         
}~,
\eea
where $\xi = -1 ( +1)$ corresponds to the left (right) handed lepton in the final state.
In the Standard Model, this polarization asymmetry comes from the
interference of the vector or magnetic moment and axial vector operators.
If in eq.(33) the lepton mass is equated to zero, our results
coincide with the results in \cite{R20} and if $m_l \neq 0$  they coincide with
the results in \cite{R11}.

\section{Numerical Analysis}

For  the input parameters which enter the sum
rules for the formfactors and the expressions of the decay width 
we have used the following values :
\bea
m_b=4.8~GeV,~~m_c=1.35~GeV,~m_\mu=0.106~GeV,~m_\tau=1.78~GeV,\nnb \\ \nnb \\
\Lambda_{QCD}=225~MeV,~~m_B=5.28~GeV,~~m_{K^*}=0.892~GeV,~~s_0=36~GeV^2,~~M^2=8~GeV^2
\nnb
\eea

In Fig.1 we present the $q^2$ dependence of the formfactors
$V(q^2),~A_1(q^2),~A_2(q^2)$ and $A_0(q^2)$ (the formfactor $A_3$ can be easily
obtained from eq.(28)). All these formfactors increase with $q^2$. From
these figures we see that $A_2(q^2)$ increases with $q^2$ strongly, but
$A_1(q^2)$ and $ A_0(q^2)$ do so smoothly. At this point let us compare our
results on these formfactors with the results which are obtained from 3-point
QCD sum rules analysis in \cite{R9}. In our case $A_1 (q^2)$ increases with $q^2$, but in
\cite{R9} it decreases with $q^2$. The behaviour of the other formfactors are
similar.

In Fig.2 we depict the dependence of the formfactors $T_1,~T_2$, and $T_3$
on $q^2$. In this case also all formfactors increase with $q^2$. For
formfactors $T_2$ and $T_3$, our predictions on their $q^2$ dependence also
differ from the predictions of \cite{R9}. In \cite{R9}, $T_2$ is positive
and smoothly decreases, the value of $T_3$ is negative for all $q^2$. Note
that our predictions on the $q^2$ dependence of all formfactors coincide
with relativistic quark model predictions \cite{R11}.
The source of discrepeancy of our results with the predictions of \cite{R9} on
$A_1,~T_2,$ and $T_3$ should be carefully analysed. This lies out of the
scope of this paper. We are planning to come back to the analysis of these
points in our forthcoming works.

In Fig.3(4)  we present the $q^2$ dependence of the
branching ratios for $B \rar K^* \mu^+ \mu^-$  
$\left( B \rar K^* \tau^+ \tau^- \right)$ decay,
with and without  the long
distance effects, respectively.

In Fig.5 we plot the longitudinal polarization asymmetry $P_L$ as
a function of $q^2$ for $B \rar K^* \mu^+ \mu^-$ 
and $B \rar K^* \tau^+ \tau^-$, with $m_t = 176~GeV$,
with and without the long distance effects. From this figure we see that
$P_L$ vanishes at the threshold due to the kinematical factor $v$ and 
that the value of $P_L$ for $B \rar K^* \mu^+ \mu^-$ 
decay varies in the region
$(- 0.5 ~\div~+ 0.5)$, when the resonance $\psi,~\psi'$ mass
region is excluded. In the $B \rar K^* \tau^+ \tau^-$
decay case, without long distance effects $P_L$
is negative for all values of $q^2$, and only in the resonance $\psi'$ mass
region $P_L$
become positive. Therefore the study of the longitudinal polarization $P_L$  
can be very
useful for understanding the relative roles of the long and short distance
contributions in the $B \rar K^* l^+ l^-$ decay.

At the end of this section let us compare our results on the Branching
ratio of the $B \rar K^* l^+ l^-$ decay with those in
\cite{R9,R11}. The value of the branching ratio is close to the results of
\cite{R9}, but about $30$ times smaller than that of \cite{R11}. In our
opinion this is due to the over estimation of the formfactors in \cite{R11}.

\section{Conclusions}

We  calculate the transition formfactors for the exclusive
$B \rar K^* l^+ l^-$($l=\mu,~\tau$) decay in the framework of the lightcone 
QCD sum rules,
and investigate the longitudinal
polarization asymmetries of the muon and tau in this decay. It is shown that
some of the formfactors in light cone and 3-point QCD sum rules have
absolutely different $q^2$ dependence. It is found that the value of the
longitudinal polarization $P_L(\mu)$ in the region $(-0.5; +0.5)$ and
$P_L(\tau)$ in $(0; -0.6)$. We also calculate the integral branching ratios
and find that they are  Br($B \rar K^* \mu^+ \mu^-$) = 
$12.06$ and Br($B \rar K^* \tau^+
\tau^-$) = $0.217$.
 
Few words about the possibility of the experimental observation of this
decay are in order. Experimentally, to observe an asymmetry $P_L$ of a decay
with the branching ratio $Br$ at the $n\sigma$ level, the required number of
events is $N = \frac{n^2}{Br\,P_L^2}$ (see \cite{R11}). For example, to 
observe the $\tau$ lepton polarization at the exclusive channel $B \rar K^*
\tau^+ \tau^-$ at the $3\sigma$ level, one  
needs at least $N = 1.66 \times 10^9~B\, \bar B$ decays.
Since in the future $B$-factories, it is expected that 
$\sim~10^9$ $B$-mesons would be created per year,
it is possible to measure the longitudinal polarization asymmetry of the
$\tau$ lepton.

\begin{figure} 
\vspace{30.0cm}
    \includegraphics{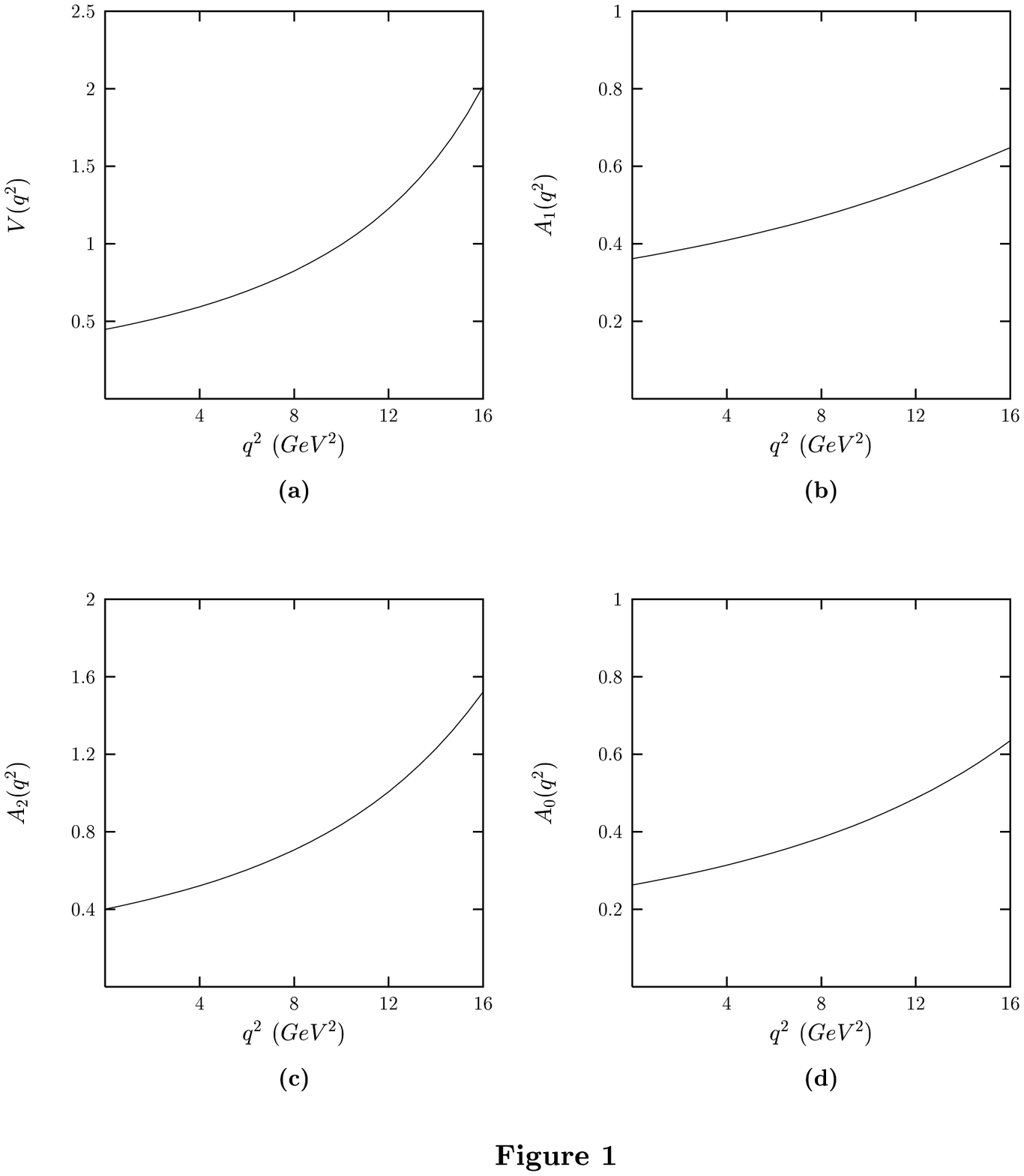}
    \vspace{-15.0cm}
\end{figure}

\begin{figure} 
\vspace{30.0cm}
    \includegraphics{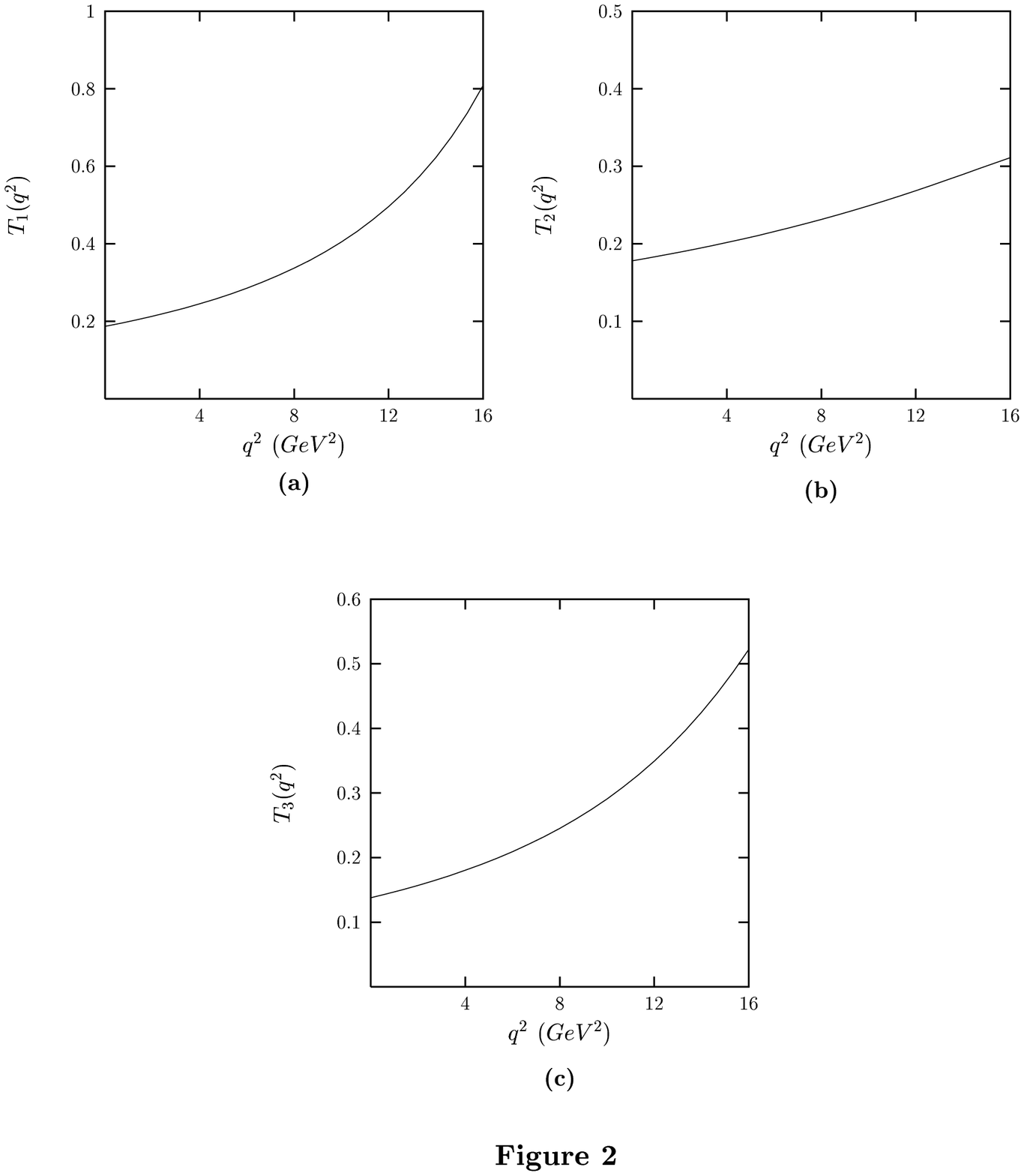} 
    \vspace{-15.0cm}
\end{figure}

\begin{figure} 
\vspace{30.0cm}
    \includegraphics{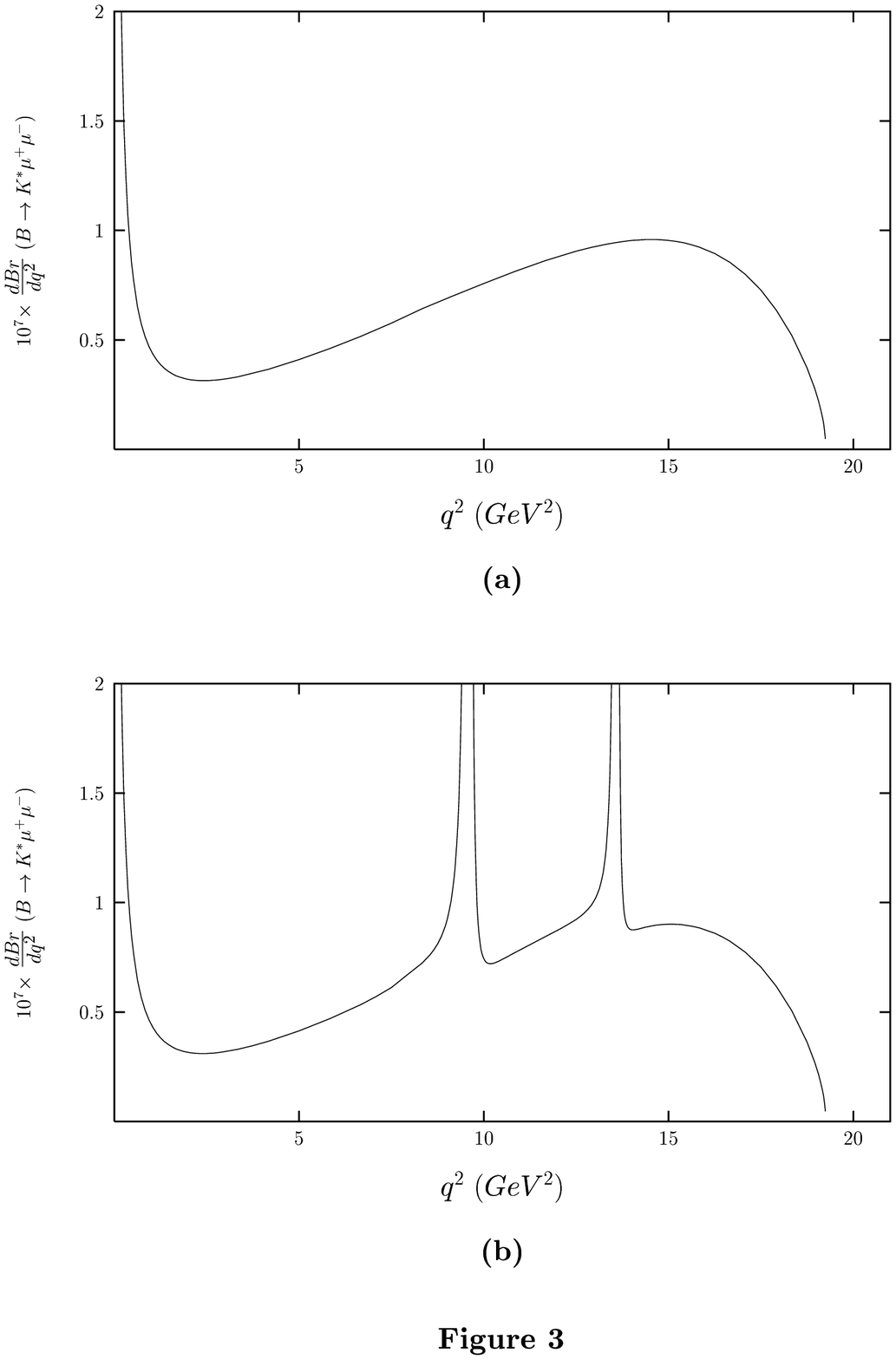} 
    \vspace{-15.0cm}
\end{figure}

\begin{figure} 
\vspace{30.0cm}
    \includegraphics{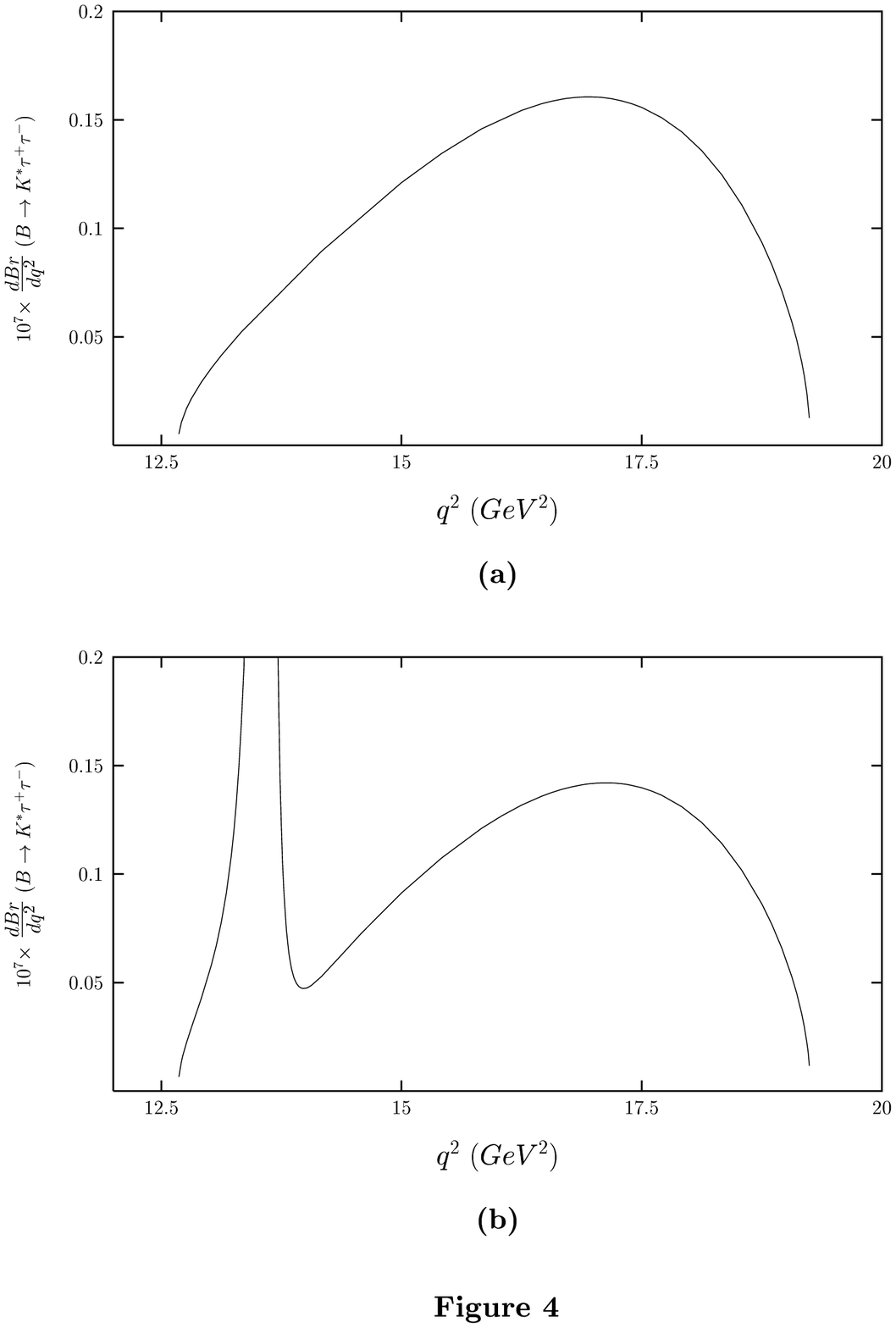}
    \vspace{-15.0cm}
\end{figure}

\begin{figure} 
\vspace{30.0cm}
    \includegraphics{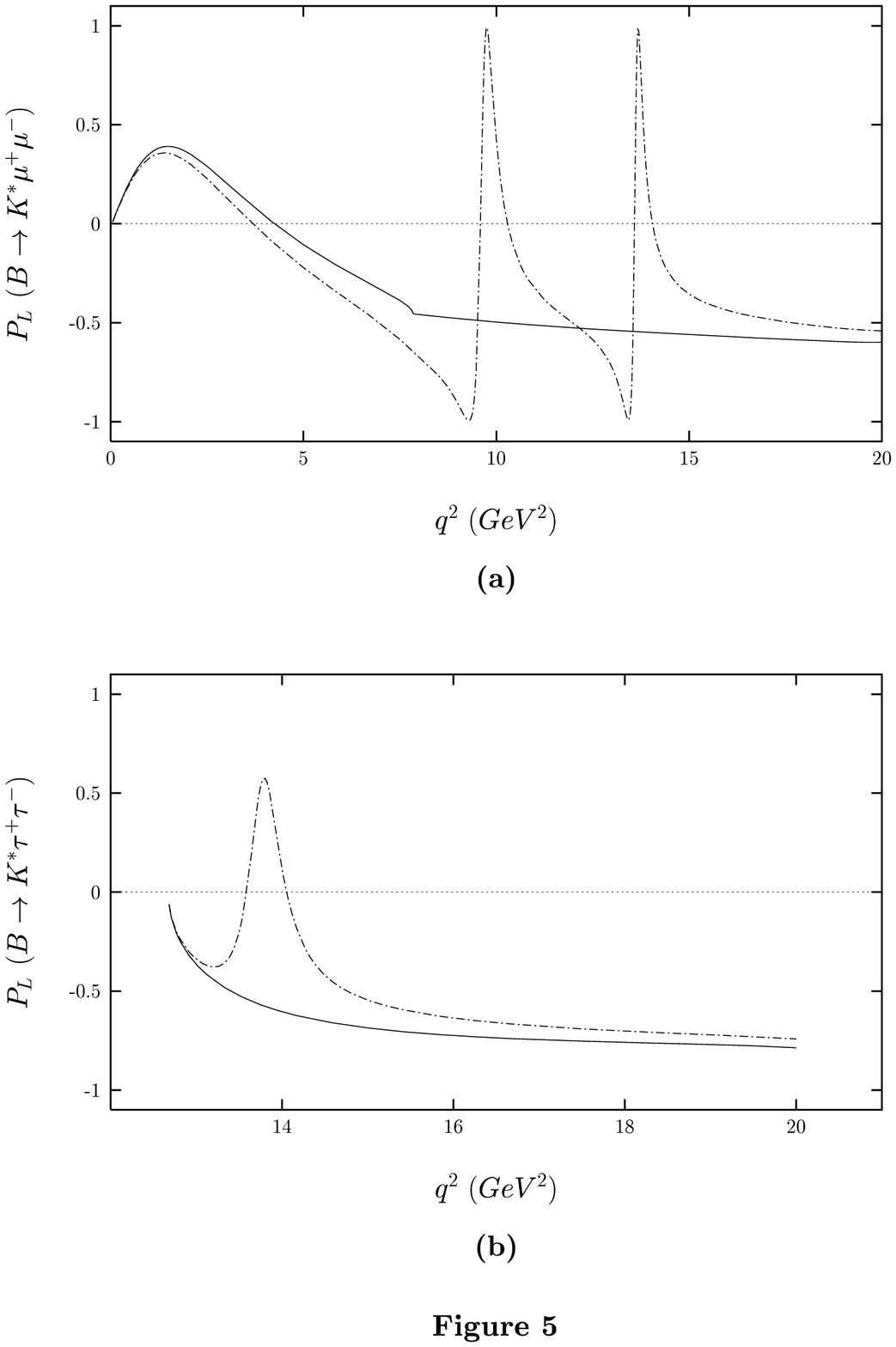}
    \vspace{-15.0cm}
\end{figure}

\newpage

\section*{Figure Captions}
{\bf 1.} The $q^2$ dependence of the formfactors $V(q^2),~A_1(q^2),~A_2(q^2)$
and $A_0(q^2)$. \\ \\
{\bf 2.} The $q^2$ dependence of the formfactors $T_1(q^2),~T_2(q^2)$ and
$T_3(q^2)$. \\ \\
{\bf 3.} a) Invariant mass squared distribution of the lepton pair for the
decay $B \rar K^* \mu^+ \mu^-$ 
which includes only the short distance contributions. \\
b) The same as in a) but including long distance effects, too. \\ \\
{\bf 4.} The same as in Fig.3, but for $B \rar K^* \tau^+ \tau^-$ decay. \\ \\
{\bf 5.} a) The longitudinal polarization asymmetry $P_L$ for the 
$B \rar K^* \mu^+ \mu^-$
decay. \\
 b) The same as in a), but for $B \rar K^* \tau^+ \tau^-$ decay. \\
In these figures, solid line corresponds to the short distance contributions
only and  dashed to the sum of both short and long distance contributions. \\
\\

\end{document}